\documentstyle[psfig,twocolumn,aps,graphicx,amsmath,array,cite,citesort,english]{revtex}
\tighten

\draft

\title{\bf High Temperature Expansion for Frustrated and Unfrustrated
$\mathbf{S\!=\!\frac{1}{2}}$ Spin Chains}

\author{Alexander B\"uhler$^1$\thanks{
e-mail: ab@thp.uni-koeln.de}
Norbert Elstner$^2$\thanks{ e-mail:
norbert@brahms.physik.uni-bonn.de} and G\"otz S. Uhrig$^1$\thanks{
e-mail: gu@thp.uni-koeln.de \hspace*{\fill} {\protect\linebreak}
internet: www.thp.uni-koeln.de/\~{}gu/}}

\address{$^1$ Institut f\"ur Theoretische Physik, Universit\"at zu
  K\"oln, Z\"ulpicher Str. 77, D-50937 K\"oln, Germany}
\address{$^2$ Physikalisches Institut, Universit\"at Bonn,
  Nu{\ss}allee 12,
D-53115 Bonn, Germany\\[1mm]
  {\rm(\today)} }

\begin{document}

\maketitle

\begin{abstract}
A computer aided high temperature expansion of the
magnetic susceptibility and the magnetic specific heat is
presented and demonstrated for frustrated and unfrustrated
spin chains. The results are analytic in nature since the calculations
are performed in the integer domain. They are
provided in the form of polynomials allowing quick and easy fits.
Various representations of the results are discussed. Combining
high temperature expansion coefficients and dispersion data yields
very good agreement already in low order of the expansion which
makes this approach very promising for the application to other
problems, for instance in higher dimensions.
\end{abstract}

\pacs{05.10.-a, 75.40.Cx, 75.10.Jm, 75.50.Ee}

\narrowtext

\section{Introduction}
Spin systems are among the most investigated systems in solid
state physics. They represent problems with high correlation since
the spin algebra does not have the simplicity of the fermionic or
the bosonic algebra. This can also be seen from the generic
derivation of antiferromagnetic spin models from a half-filled
Hubbard model in the limit of large interaction $U\to \infty$.
Hence even the calculation of simple properties like the magnetic
susceptibility $\chi$ or the magnetic specific heat $C$ is not
straightforward. Experimentally, however, susceptibility and
specific heat are the first quantities used to characterise a
compound. So quantitative theoretical predictions are very important to
pinpoint the appropriate model.

Quantum Monte Carlo methods underwent considerable progress in the
last years so that the calculation of $\chi$ and $C$ for
unfrustrated systems has become feasible to very high accuracy.
The treatment of frustrated spin systems, however, is still a
difficult task since in the standard Ising basis the sign problem
occurs. This leads to the phenomenon that considerable
cancellations and the concomitant loss of statistical accuracy
occur in the quantum Monte Carlo computation, see  e.g.\
Ref.~\cite{johns00a}. In the case of strong frustration one still
has to resort to complete diagonalisation which restrict the
accessible system sizes very much, see e.g. Ref.~\cite{fabri97a}. For
(quasi) one dimensional systems like chains or ladders finite
temperature density-matrix renormalisation provides a reliable
means to calculate susceptibilities and specific heats, see e.g.\
Ref.~\cite{bursi95,shiba97,wang98,johns00b}. The caveat, however,
remains that the  numerical methods require a new programme run
for each set of parameters. So fitting becomes a tedious task as
soon as more than one parameter is involved.

The objective of the present article is to introduce a computer
aided expansion in the inverse temperature for the quantities
$\chi$ and $C$. This method is a variant of the ``linked cluster''
approach \cite{he90,gelfa90}. In the form implemented here it
provides the results as polynomials in the relevant energy ratios.
Thereby extremely fast and convenient fit procedures become
possible. Frustrating terms do not pose more problems than any
other additional couplings. For the sake of simplicity we will
demonstrate our approach for one dimensional systems, i.e. chains.
The subsequent choice of an appropriate representation of the results is
 also very important. The inclusion of low temperature information enlarges 
the range of validity considerably.

In the next two sections the method is explained in detail and
contrasted to the conventional linked cluster approach. In
Sect.~\ref{sec:results} the results are given and represented in
various ways in order to obtain the best description. A particular
representation based on additional dispersion data is explained in
Sect.~\ref{sec:dispers}. The findings are summarised in the
concluding Sect.~\ref{sec:conclusion}.

\section{Methods}
\label{sec:methods} In the present work two methods are used to
expand the physical quantities. The first one is the linked
cluster method \cite{he90,gelfa90} and the second one a method
which will be called \textit{moment algorithm} henceforth. To be
explicit, the spin-$\frac{1}{2}$ Heisenberg chain
\begin{equation}
  \label{eq:heisenberg}
   H=\sum_{i=1}^N(J\vec{S}_i\vec{S}_{i+1}+
    \alpha J \vec{S}_i\vec{S}_{i+2}-h{S}_i^z)\
\end{equation}
with nearest and next nearest neighbour interaction  is
investigated with $h$ representing the magnetic field in units of
$g\mu_B$. The ratio of nearest and next nearest neighbour exchange
coupling is given by $\alpha$.

The main idea of the linked cluster method is to restrict
calculations to finite systems to obtain results in the
thermodynamic limit. The method can be applied to systems
(clusters) which are described by a sum of local hamiltonians like
the Heisenberg hamiltonian in (\ref{eq:heisenberg}). An expansion
of a quantity in powers of such a hamiltonian results in the
computation of the contributions of clusters of various sizes. To
obtain the contribution of a finite cluster to the quantity of the
infinite system all contributions of subclusters have to be
subtracted with suitable multiplicity. Only connected (``linked'')
clusters provide nonvanishing contributions. The non-connected
clusters cancel out due to the normalisation of the expectation
value (see for instance Eq.~(\ref{eq:chitrace})). 
This is the main result of the linked cluster  theorem.

The numerical approach of the moment algorithm makes use of the
result of the linked cluster theorem. Here as well the physical
quantities are evaluated in the thermodynamic limit by means of
finite systems. Let us consider the Heisenberg hamiltonian with
nearest neighbour interaction acting on a chain and let us expand
a physical quantity in powers of $\beta J=J/T$, i.e.\ in powers of
this hamiltonian. The largest connected cluster at a given order
$n$ of the expansion contains $(n+1)$ sites. Hence all clusters of
this size must be embedded completely in the finite system to
obtain valid results in the thermodynamic limit.

For concreteness, we compute the magnetic susceptibility per site
at vanishing magnetic field
\begin{equation}
  \label{eq:chitrace}
  \chi (T) = \frac{\beta}{N}\frac{{\text{tr}}(M^2 e^{-\beta
  H})}{{\text{tr}}(e^{-\beta H})}\ .
\end{equation}
Denominator and numerator are computed separately by expanding the
corresponding exponential functions. The resulting rational
function is again expanded around $\beta=0$ to obtain
  a polynomial in the inverse temperature $\beta$.

On this stage the comparison of the moment algorithm to the linked
cluster approach shows one advantage and one disadvantage. The
advantage is that it is not necessary to determine and to classify
all contributing clusters explicitly. This task may not be
underestimated in view of the lack of efficient algorithms
comparing graphs. This point matters in particular for complicated
lattices with different types of bonds. In this respect, the
moment algorithm is simpler than the linked cluster approach. The
disadvantage is that the finite systems which have to be dealt
with are fairly large, in particular for elevated orders in
$\beta$ and higher dimensions. 
In this respect, the linked cluster approach is better
suited for higher dimensional problems.

The disadvantage mentioned is less troublesome if an efficient way
to compute traces of powers of the hamiltonian is available. To
this end we present an algorithm which computes such traces in
a very fast way. ``Fast'' means that the necessary effort increases
not exponentially with system size $N$ but only in powers of $N$.
The algorithm reduces the trace to an ordinary expectation value
in a higher-dimensional Hilbert space. To this purpose the Hilbert
space of the real system is doubled by introducing to each real
site $|i_r\rangle$ a ``doubled'' site $|i_d\rangle$. Any operator
$A$ defined on the real Hilbert space acts on the tensor product
of real and doubled Hilbert space in the canonical way. That is
$A$ becomes $A\otimes {\boldmath 1}$ acting as the identity on the
doubled Hilbert space.

 Furthermore, we consider a state which is the (unnormalised) product
of singlets (or triplets with $S^z=0$) between the  real and the
doubled sites
\begin{equation}
  \label{eq:singulett}
  |S\rangle = \prod_{i=1}^N (|\uparrow_r\downarrow_d\rangle
                -|\downarrow_r\uparrow_d\rangle)\big|_i\ .
\end{equation}
The key observation is that the trace in the original, real
Hilbert space is identical to the expectation value of $|S\rangle$
in the extended Hilbert space
\begin{equation}
  \label{eq:trace}
  {\text{tr}}(A)|_{\text{real}} = \langle S|A\otimes {\boldmath 1}
|S\rangle\  |_{\text{double}}\ .
\end{equation}
To see the identity (\ref{eq:trace}) one considers first a single
site. Explicit calculation shows
\begin{mathletters}
  \label{eq:singlespin}
\begin{eqnarray}
  {\text{tr}}(A)|_{\text{real}} &=& \langle \uparrow |A| \uparrow
  \rangle + \langle \downarrow |A| \downarrow \rangle \\
  &=&\langle \uparrow_r\downarrow_d|A|\uparrow_r\downarrow_d\rangle
  |_d + \langle
  \downarrow_r\uparrow_d|A|\downarrow_r\uparrow_d\rangle|_d \\
  &=&\langle S|A|S\rangle|_{d}
\end{eqnarray}
\end{mathletters}
where we used the subscripts $r$ and $d$ for `real' and `doubled'
(extended) Hilbert space, respectively. The validity of
(\ref{eq:trace}) follows from (\ref{eq:singlespin}) directly  for
all operators $A$ which are products of local spin operators since
the Hilbert space of many spins is just the tensor product of the
Hilbert spaces of the individual spins
\begin{mathletters}
\begin{eqnarray}
  \label{eq:product}
&&  {\text{tr}}\prod_iA_i|_r = \prod_i {\text{tr}}A_i|_r\\ &&=
\prod_i \langle(\langle\uparrow_r\downarrow_d|
                -\langle\downarrow_r\uparrow_d|)
                A_i(|\uparrow_r\downarrow_d\rangle
                -|\downarrow_r\uparrow_d\rangle)|_{i}\\
 &&= \langle S|\prod_iA_i|S\rangle|_{d}\ .
\end{eqnarray}
\end{mathletters}
>From the linearity of the expectation value and the trace follows
then that (\ref{eq:trace}) holds also for all operators $A$ since
they can be decomposed into sums of products of local spin
operators.

On the left hand side of (\ref{eq:trace}) for $A=H^2$ for instance
one has to compute in the Ising basis (spins up or down) $2^N$
contributions for $L^2$ terms ($N$ number of sites, $L$ number of bonds),
 i.e.\ one has to sum about $L^2
2^N$ terms. On the right hand side of (\ref{eq:trace}), however,
one starts with a single state which will be excited in $L$ ways
by the application of $H\otimes \boldmath1$ and requires to be
de-excited in the same way so that only $L$ terms contribute in the
end. Thus one saves an exponential factor.
An obvious by-product are the relations
\begin{mathletters}
\label{eq:relations}
\begin{eqnarray}
\nonumber \langle S|H^{2n}|S\rangle &=&\langle S|H^{n}
H^{n}|S\rangle \\ &=&  \left|H^{n}|S\rangle\right|^2 \\ \langle
S|H^{2n+1}|S\rangle &=&\langle S|H^{n}   H^{n+1}|S\rangle
\end{eqnarray}
\end{mathletters}
which imply that for a given order $m$ in $\beta$ one needs to
calculate only about $m/2$ applications of $H$ to the singlet
product state $|S\rangle$. This statement remains true for the
numerator of susceptibilities as in (\ref{eq:chitrace}) if the
observable (here: $M=\sum_{i=1}^N S^z_i$) commutes with $H$. 
This is the case for the
uniform magnetisation $M$. Replacing
 $|S\rangle $ by  $M|S\rangle $ thus makes the relations (\ref{eq:relations})
also applicable to the numerator of (\ref{eq:chitrace}).

We implemented the actual calculations on computer. For not too
high orders this can still be done with computer algebra
programmes. But to obtain the highest orders it is necessary to
write task-specific programmes. This was done by using the
language C$^{++}$. Yet the dependencies on all coupling constants
are included on the symbolic level, i.e.\ in polynomials of the
coupling constants. So, once obtained, the results are available
to everybody and they can be fitted to any experimental curve very
quickly and easily. The results for the moment algorithm
(frustrated (order 10) and unfrustrated spin chain (order 16)) and
for
 the linked cluster algorithm (unfrustrated spin chain only, but order 24)
are presented in the appendix.

Before concluding this section we like to note that the trick to
pass from a trace to an expectation value in a higher dimensional
Hilbert space is not
 restricted to spin-$\frac{1}{2}$ systems. By introducing for instance
the generalised product state of singlets $|S\rangle_g$
\begin{equation}
  \label{eq:singulettgen}
  |S\rangle_{g} = \prod_j\sum_{i=0}^{2S}
    \frac{(-1)^i}{\sqrt{2S+1}}|(S-i)_r,(i-S)_d\rangle\Big|_j
\end{equation}
the method can be applied to arbitrary spin. In
(\ref{eq:singulettgen}) $|j_r,l_d\rangle$ stands for the state
where the real spin has $S^z=j$ and the doubled spin $S^z=l$. For
a derivation one simply has to redo the calculation
(\ref{eq:singlespin}).

Henceforth the angular brackets will denote the expectation value
with regard to $|S\rangle$ or the original trace, respectively,
since their identity is established and so no further
distinction necessary.

The calculations in the present work were done to the highest
order possible on the available work stations. For the
unfrustrated chain the physical quantities were expanded to order
$N$ on a finite system of size $N$ sites. Therefore clusters in
the $N$th order are overcounted or missed. But it is possible
to correct these effects in highest order by an analytical argument
which is presented in the subsequent section.

\section{Finite size corrections \label{sec:finitesize}}
The wrap-around effects of the numerator and denominator of
Eq.~(\ref{eq:chitrace}) are investigated separately for a chain of
length $N$ with periodic boundary conditions.

The denominator has the following kind of contributions in the
$N$th order
\begin{equation}
  \label{eq:wrapden}
  \langle H^N \rangle = \langle \prod_{i=1}^{N}
(\vec{S}_{i}\vec{S}_{i+1})\rangle + \ldots
\end{equation}
which are {\em not} realised in the thermodynamic system. Fixing
the component of {\em one} of the spin vectors involved to for
instance $S^x$ one sees that a non-vanishing contribution occurs
only if {\em all} spin components are $S^x$. The overall value of
the right hand side in (\ref{eq:wrapden}) is $4^{-N}$. Since all
permutations of the scalar products will occur as well if the
left hand side of (\ref{eq:wrapden}) is expanded and since all
these permutations yield the same contribution the factor $N!$
has to be added. Since the choice of the spin component was
arbitrary an additional factor 3 concludes the argument. Thus one
has to subtract
 $3N!4^{-N}$ to yield the thermodynamic result in the denominator.

The corrections of the numerator of Eq.~(\ref{eq:chitrace}) are
more complicated. They consist of three contributions. The first
is similar to the one in the denominator
\begin{equation}
  \label{eq:wrapnum1}
  \langle (S_1^z)^2 H^N\rangle = \langle(S_1^z)^2
  \prod_{i=1}^{N}(\vec{S}_i\vec{S}_{i+1} \rangle + \ldots
\end{equation}
and overcounts the numerator by $3N!4^{-N-1}$. On the other hand,
the thermodynamic contribution is neglected in this order of expansion
\begin{equation}
  \label{eq:wrapnum2}
  \langle S_1^z S_{N+1}^z(\vec{S}_1\vec{S}_2)(\vec{S}_2\vec{S}_3)
  \ldots(\vec{S}_N\vec{S}_{N+1})\rangle
\end{equation}
which represents the second correction. It takes the value
$2N!4^{-N-1}$. The factor $N!4^{-N-1}$ arises for the same reasons
as before. There is no factor 3 since the spin component is
already fixed by the choice of the magnetisation direction. The
geometric factor 2 arises instead because the sites 2 to $N+1$ can be found
to the right or to the left of the starting site 1.

The third and last correction consists of clusters with triple
occurrence of two sites. It is based on the identity for 
$S=1/2$
\begin{equation}
\label{eq:identity} S^x S^y S^z  = i/8
\end{equation}
which holds also for all cyclic permutations of the spin
components. For anticyclic permutations the left hand side of
(\ref{eq:identity}) acquires a minus sign.

A wrap-around with sites occurring three times is possible as soon
as the magnetisation operators are taken to be at different sites:
$S^z_1 S^z_j$ with $j\neq 1$. Then each of the sites 1 and j
appear three times
\begin{equation}
\label{eq:three} \langle S^z_1 S^z_j \prod_{i=1}^N
(\vec{S}_i\vec{S}_{i+1}\rangle\ .
\end{equation}
Note that all permutations of the sequence of the scalar products
appear. Hence in most cases the contributions cancel each other
since cyclic and anticyclic permutations of the spin components at
site 1 or site $j$ occur independently and with equal amplitude.
 Only if the site $j$ is adjacent to
1, i.e.\ $j=2$ or $j=N$ (periodic boundary conditions),  the
cyclic and anticyclic permutations at site 1 and $j$ are
correlated and a finite total effect remains. The relevant factors
are (for $j=2$)
\begin{equation}
  \label{eq:wrapnum3}
  \langle S_1^z S_2^z(\vec{S}_N \vec{S}_1)(\vec{S}_1 \vec{S}_2)
  (\vec{S}_2 \vec{S}_{3}) \ldots \rangle\ .
\end{equation}
Among the $3!=6$ ways to arrange the three scalar products in
(\ref{eq:wrapnum3}) the two where $(\vec{S}_1\vec{S}_2)$ is in the
middle yield $1/8^2 (S_N^xS_3^x+S_N^yS_3^y)$
 while the other four yield $-1/8^2(S_N^xS_3^x+S_N^yS_3^y)$
so that $-2/8^2 (S_N^xS_3^x+S_N^yS_3^y)$ remains. Accounting for
the multiplicity due to the arrangement of the other scalar
products yields the combinatorial factor $N!/3!$. A factor 2 comes
from the possibility to choose $j=2$ or $j=N$. The overall third
correction finally reads $-8\cdot 4^{-N+1} N!/3!$.

In summary, the total corrections to the results computed in the
$N$th order for a finite system of $N$ sites with periodic
boundary conditions for the numerator Nu and for the denominator
De   are
\begin{mathletters}
  \label{eq:wrapsum}
\begin{eqnarray}
  {\text{Nu}}_{\text{corrected}} &=&
  {\text{Nu}}_{\text{computed}} + N!\cdot\frac{1}{3}
  \left(\frac{1}{4}\right)^{N+1} \\
  {\text{De}}_{\text{corrected}} &=&
  {\text{De}}_{\text{computed}} - N!\cdot 3
  \left(\frac{1}{4}\right)^{N} \ .
\end{eqnarray}
\end{mathletters}

After these considerations it is also straightforward to correct
the wrap-around effects for the Heisenberg chain with next-nearest
neighbour interaction. In the $N$th order on a finite system of
$2N$ sites these effects occur only in the $N$th order in $\alpha$
and in $\beta$. In this order the system
corresponds to a system of two independent chains with  nearest
neighbour interactions only, so that the corrections
(\ref{eq:wrapsum}) apply to terms of $N$th order in $\alpha$. This
concludes the discussion of the finite size corrections.

\section{Results and Representations}
\label{sec:results} In the appendix the series coefficients are
presented for the magnetic susceptibility $\chi$ and for the
magnetic specific heat $C$ per site. The specific heat is derived
from the denominator of Eq.~(\ref{eq:chitrace}) which is the
partition function of the system by
\begin{mathletters}
\begin{eqnarray}
  \label{eq:heat1}
  C(T)&=& \frac{1}{N}\frac{\partial}{\partial T}
\frac{\langle H e^{-\beta H}\rangle}{\langle e^{-\beta H}\rangle}
\\
&=&\frac{1}{N}\frac{\partial}{\partial T} \left(
  \frac{-\frac{\partial}{\partial \beta}\langle e^{-\beta H}\rangle
  }{\langle e^{-\beta H}\rangle} \right)\ .
\label{eq:heat2}
\end{eqnarray}
\end{mathletters}
It is worth mentioning that due to the derivation in
(\ref{eq:heat2}) one order in $\beta$ is lost. It is re-gained,
however, by the subsequent derivation with respect to $T$.

In particular, results for the unfrustrated chain are listed in
Appendix 3. These are obtained by the linked cluster method and
comprise orders as high as 24. Such high orders are obtained
by an exact extrapolation of smaller clusters as explained below.
 They are in agreement with expansion
results previously obtained \cite{baker64,oboka67}. 
In Appendix 3 a the magnetic
susceptibility coefficients are given, in Appendix 3 b the specific
heat coefficients are given. 

For the loop-free chain the calculation of high temperature series for 
spin models simplifies decisively compared to higher dimensions due to the 
simple structure of the contributing cluster.
A finite open chain with $n$ bonds will contribute to the specific heat
 only in order $\beta^{2n}$. This is so because otherwise there will always be 
 one site occuring an odd number of times in the spin products leading
eventually to a vanishing trace.

The susceptibility series shows a systematic pattern which can be exploited
to obtain longer series. 
For the $S=1/2$ Heisenberg chain, the nonvanishing contribution
of order $\beta^n$ of a finite open cluster with $n$ bonds and 
$n+1$ sites  will be due to one spin product of the hamiltonian acting on 
each bond and two magnetisation operators $S^z$ at the ends of the chain. 
Its contribution is of the form $ a_0 n!\, 4^{-n-1} \beta^n$,
where $a_0$ is independent of the  cluster size (as before we consider the
terms without the $1/n!$ factors from the exponential series).

Similarly, the contribution in order $\beta^{n+1}$ will be of the form
$ (b_0 + n b_1) n!\, 4^{-n-2} \beta^{n+1}$.
Again $b_0$ and $b_1$ do not depend on the length of the finite cluster.
The term proportional to $n$ is due to the $n$ possibilities to 
attach one more term of the hamiltonian to any of the $n$ bonds. 

In the same way, the general form of the higher order contributions can be
determined with more and more coefficients. 
With sufficiently large clusters the coefficients $a_0$, $(b_0, b_1)$, 
$(c_0, c_1, c_2)$, $\dots$  can be obtained by solving a system of linear 
equations. Using clusters 
with up to 18 bonds allowed to extend the susceptibility series to order 24.

The moment algorithm allowed us to compute results for the
unfrustrated chain up to order 16 for the susceptibility
(\ref{sec:momentsusiNN}) and for the specific heat
(\ref{sec:momentcNN}). Note that this is only two orders less than the 
maximum cluster which is actually computed in the linked cluster approach.
The results are in complete agreement with
the linked cluster results. The only difference is in the
internal representation where doubles are used in the linked
cluster programme and true fractions in the moment algorithm. The
susceptibility as well as the specific heat of the frustrated
chain is expanded up to order 10 (\ref{sec:momentsusiNNN}  and
\ref{sec:momentcNNN}).

Having obtained the series for the various expansions we pass now
to the discussion of suitable representations. The choice of an
appropriate representation allows to gain the maximum of
information from the bare series coefficients. We follow two main
routes. One is the use of Pad\'e approximants and continued
fractions, respectively; the other is to incorporate additional
information at low temperatures to improve the representations in
the low temperature regime.

Exact results from Bethe ansatz calculations for the unfrustrated
chain \cite{klump93b,klump98} and numerical results from
density-matrix renormalisation group (DMRG) calculations for the
frustrated chain \cite{klump99a} are used as benchmarks.

\subsection{Unfrustrated Chain}

Since the convergence of the plain series in $\beta$ can be
hindered by any pole the use of a Pad\'e approximant describing
the quantity under study by a rational function is more stable than
the plain series.
\begin{figure}[htbp]
  \begin{center}
    \includegraphics[width=\columnwidth]{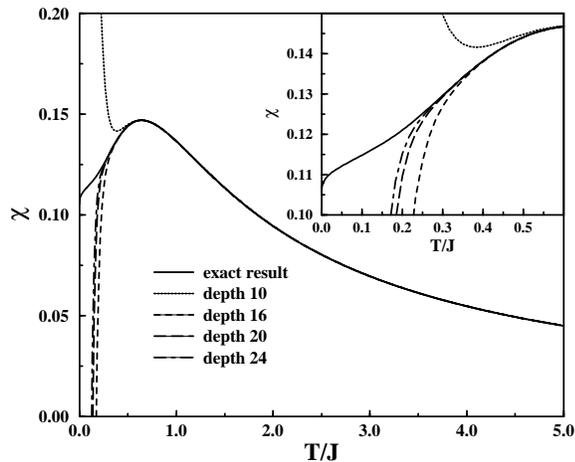}
    \caption{Comparison of various depths of continued fraction
      representations of the susceptibility $\chi$ for
      the unfrustrated chain. }
    \label{fig:chi_comp_pades}
  \end{center}
\end{figure}
The polynomial in $\beta$ of the physical quantity under
consideration is represented as a continued fraction, say with
depth $2N$,
\begin{equation}
  \label{eq:contfrac}
  \chi[2N](\beta)=
    \cfrac{\beta}{c_1+
      \cfrac{\beta}{c_2+
        \cfrac{\ddots}{\cdots +
          \cfrac{\beta}{c_{2N}}}}}
\end{equation}
which is equivalent to the $[N,N]$ Pad\'e approximant. An odd
depth of $2N\!+\!1$ is equivalent to the $[N+1,N]$ Pad\'e
approximant. Increasing the degree of either the numerator or the
denominator polynomial at the expense of the other does not
improve the results. The advantage of the continued fraction
representation is that the coefficients $c_i$ remain constant on
increasing depth. For instance $c_1$ is equal to $4$ and $c_2$
is equal to $1/2$ due to the Curie and due to the Curie-Weiss law,
respectively.

In Fig.~\ref{fig:chi_comp_pades} various depths of continued
fraction representations of the unfrustrated susceptibility are
shown. The comparison with the exact result shows that the
agreement improves on increasing depth as expected. But the
improvement of the representations is relatively
small for higher depths. Excellent agreement can be achieved down to
 $T\approx J/4$ if coefficients up to order 24 are used.

In order to extend the region of satisfying agreement to lower
temperatures information about the low temperature regime can be
incorporated. To this end, the continued fraction depth is
incremented by adding a new constant. This constant is not
determined from the high temperature expansion. But it is
determined such that the desired additional property is fulfilled.

To be precise, we include the value of the susceptibility at zero
temperature. For the unfrustrated chain it can be expressed as
\cite{mulle81,klump93b}
\begin{equation}
  \label{eq:chi0}
  \chi(0) = \frac{1}{2\pi}\frac{1}{v_S}
\end{equation}
with the spin wave velocity $v_S=\frac{\pi}{2}$ \cite{cloiz62}.
Eq.~(\ref{eq:chi0}) implies that the central charge $c$ is one.
Assuming that the central charge does not vary on switching on the
frustration we will use (\ref{eq:chi0}) there, too. For the change
of the spin wave velocity is accounted by \cite{fledd97} by
\begin{equation}
  \label{eq:chi0frust}
  v_S = \frac{\pi}{2}(1-1.12\alpha)\text{ for }0\le\alpha<\alpha_c\ .
\end{equation}
In the gapped regime for $\alpha\ge\alpha_c\approx 0.241167$ \cite{egger96}
 the susceptibility
vanishes exponentially at $T=0$. 

The relevant gap, however,
is not the spectroscopic gap $\Delta_{01}$ between the $S=0$ ground state and
$S=1$ excited states but half of this value $\Delta_{01}/2$. This is so since
the elementary excitations of strongly frustrated spin chains are 
asymptotically free massive 
$S=1/2$ spinons, see for instance Refs.~\cite{shast81a,affle97,uhrig99a}.

The low temperature behaviour of the specific heat \cite{klump93b}
is given by
\begin{equation}
  \label{eq:heat0}
  C(T\approx 0 ) = \frac{\pi}{3}\frac{1}{v_S}\cdot T
\end{equation}
with the same spin wave velocities $v_S$ as in the previous
equations 

for $\alpha<\alpha_c$. In the gapped regime for supercritical
frustration, an exponential vanishing for low temperatures
is to be expected.
 From (\ref{eq:heat0}) follows directly
\begin{equation}
  \label{eq:heatprime0}
  \frac{d}{dT}C(T=0) = \frac{\pi}{3}\frac{1}{v_S}
\end{equation}
which can also be incorporated in the representations. A third
piece information is obtained by
  \begin{equation}
    \label{eq:heatint0}
    s(\infty)-s(0)=\int_0^{\infty}\frac{C(T)}{T}d\!T = \ln 2\ .
  \end{equation}
This piece of information, however, is more difficult to build-in
since it involves an integration over the continued fraction. Moreover,
it turns out that its effect is not sizable. Thus it is not considered
any further.

\begin{figure}[b]
  \begin{center}
    \includegraphics[width=\columnwidth]{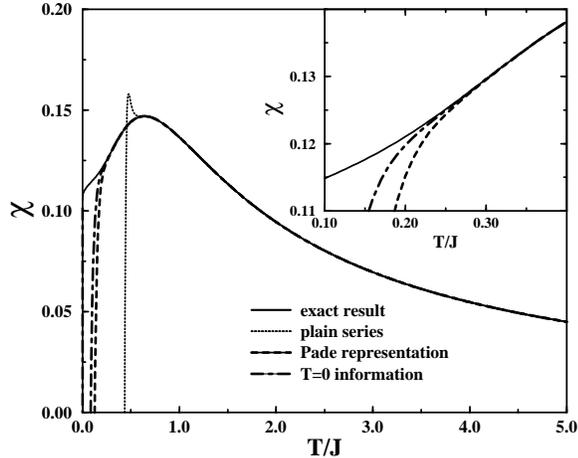}
    \caption{Various representations of the susceptibility $\chi$ for
      the unfrustrated chain. The plain series is expanded up to order
      24. The inset shows a zoom of the [12,11]-Pad\'e
      approximant and of the [12,12]-Pad\'e approximant with $T\!=\!0$
      information, see Eq.~(\ref{eq:chi0}).}\label{fig:chi_linked}
  \end{center}
\end{figure}
Fig.~\ref{fig:chi_linked} shows the various representations for
the susceptibility $\chi$ of the unfrustrated chain. The approximate results 
agree very well with the exact ones down to
$T/J\approx 0.2$ for the Pad\'e approximant with $T=0$ information.
 Without the aid of the exact result one is also able
to determine the quality of the representation by
comparison of results in highest order with those in lower orders.

\begin{figure}[t]
  \begin{center}
    \includegraphics[width=\columnwidth]{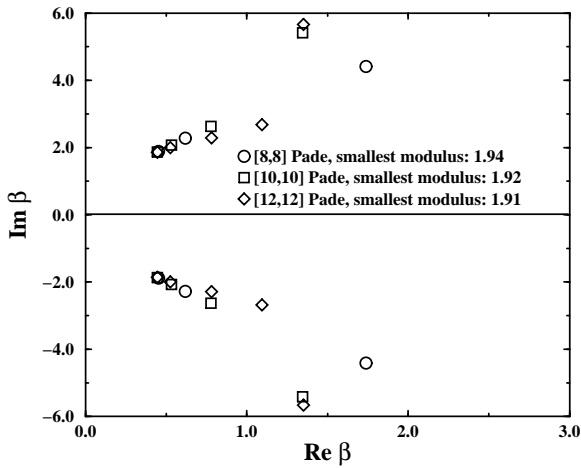}
    \caption{Singularities of various Pad\'e representations in
      $\beta$ for $\chi$ and the smallest moduli of their $\beta$ values.
      Singularities in the left half-plane are not shown.
      }\label{fig:singularities}
  \end{center}
\end{figure}
It is instructive to look at the poles of the Pad\'e representations
closest to the origin. Their modulus is an estimate for the 
radius of convergence of the plain series. From the values given in
Fig.~\ref{fig:singularities} one can deduce that this value is fairly
constant at about $\beta_{\rm max}\approx 1.9$. 
This implies that the plain series
will always diverge below about $T\approx 0.53$ irrespectively of
the order of the series, cf.~Figs.~\ref{fig:chi_comp_pades},
\ref{fig:chi_linked}. This is a good illustration of the utility of
Pad\'e representations. They are not blocked by the occurrence of poles.
So they are able to represent more complicated functional dependencies.

\begin{figure}[b]
  \begin{center}
    \includegraphics[width=\columnwidth]{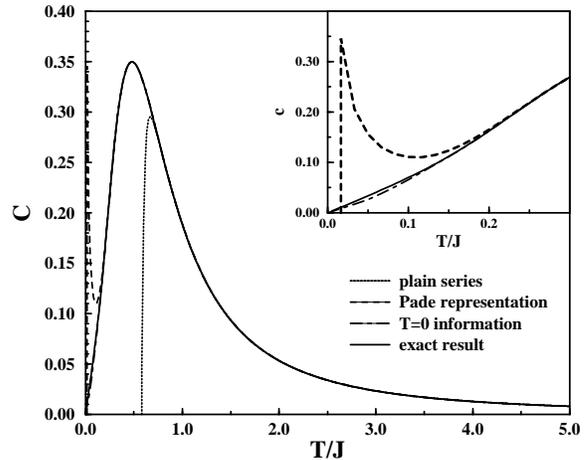}
    \caption{Various representations of the specific heat $C$ for
      the unfrustrated chain. The plain series is expanded up to order
      24. The inset shows a zoom of the [11,11]-Pad\'e
      approximant and of the [12,12]-Pad\'e approximant with $T\!=\!0$
      information, see Eq.~(\ref{eq:heat0}) and (\ref{eq:heatprime0}).
      }\label{fig:heat_moment}
  \end{center}
\end{figure}
The low temperature behaviour of the specific heat is less complex
than the one of the susceptibility. Fig.~\ref{fig:heat_moment}
shows a better agreement with the exact result, especially in the
low temperature regime. Here the plain series is expanded up to
order 24 in $\beta$. With the two pieces of low temperature
information (\ref{eq:heat0},\ref{eq:heatprime0}) the representation agrees 
very well with the exact result down to $T/J\approx 0.1$.

At this stage a consideration of the range of validity that one 
could expect is in order. In fact, we argue that one should have
expected an even better description based on $1/T$ results. 

Calculating up to order $n$ means that the physics on a length scale
$n$ (lattice constant set to unity)
 is taken into account since this is the size of the maximum cluster
treated properly. So one is led to the estimate
\begin{equation}
\label{eq:estimate}
n \approx \frac{v_S}{2\pi T_{\rm min}}
\end{equation}
where the energy scale $2\pi T_{\rm min}$ results from the discretisation of
the Matsubara frequencies which serves here as infrared cutoff.
From (\ref{eq:estimate}) follows for the unfrustrated chain $T_{\rm min} 
\approx 1/(4 n)$ for the temperature down to which the large $T$ 
information should be capable to describe the physics properly.
It is obvious that the validity stops actually at much higher
temperatures. For this reason we presume that the representation
by a Pad\'e approximant is not yet the optimum.

\subsection{Frustrated Chain}

Motivated by the inorganic spin-Peierls system
$\text{CuGeO}_{\text{3}}$ \cite{hase93a,bouch96}
the results for the frustrated chain are
presented with a fixed $\alpha$-value of $0.35$
\cite{riera95,casti95,fabri98a}. This value is chosen
since it allows a good description of the susceptibility data.
At low temperatures there is evidence that the frustration
is lower \cite{knett00b}.

Comparisons to benchmark calculations were also performed
at the critical frustration $\alpha_c$.
Compared to  the higher orders reached in
the unfrustrated case, the results for the frustrated chain should
agree well only for higher values of $T/J$. We will see, however,
that this is not the case.

\begin{figure}[t]
  \begin{center}
    \includegraphics[width=\columnwidth]{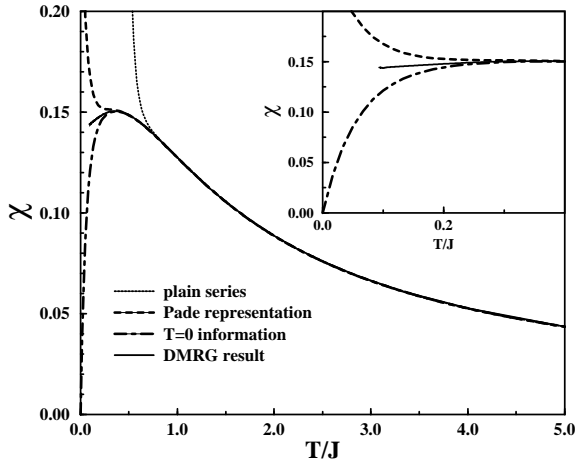}
    \caption{Various representations of the susceptibility $\chi$ for
      the frustrated chain with $\alpha=0.35$. The plain series is
      expanded up to order 10. The inset shows a zoom of the
      [5,4]-Pad\'e approximant and of the [5,5]-Pad\'e approximant with
      $T\!=\!0$ information for the gapped
      regime.}\label{fig:chifrust_moment}
  \end{center}
\end{figure}
Fig.~\ref{fig:chifrust_moment} shows the susceptibility compared
to DMRG calculations. The best representation with $T=0$
information is in very good agreement down to $T/J\approx 0.25$.

\begin{figure}[b]
  \begin{center}
    \includegraphics[width=\columnwidth]{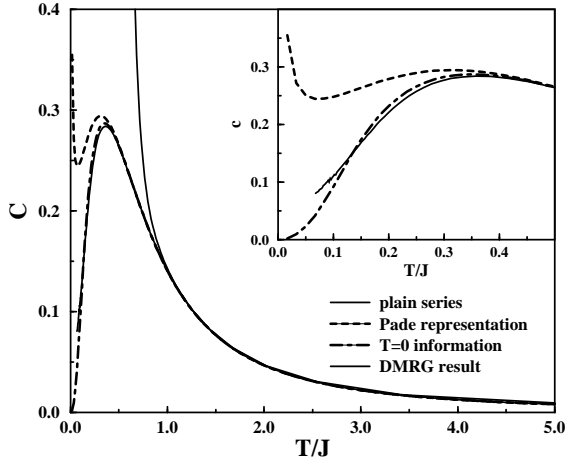}
    \caption{Various representations of the specific heat $C$ for
      a frustrated chain with $\alpha=0.35$. The plain series is
      expanded up to order 10. The inset shows a zoom of the
      [4,4]-Pad\'e approximant and of the [5,5]-Pad\'e approximant with
      $T\!=\!0$ information, see Eqs.~(\ref{eq:heat0}) and
      (\ref{eq:heatprime0})\label{fig:heatfrust_moment}}
  \end{center}
\end{figure}

Since the frustration is supercritical $0.35>\alpha_c$ the
$T=0$ information consists in fixing $\chi(T=0)=0$ due to the exponential
vanishing.
The region of satisfying agreement coincides very well with the one
of the representation of the specific heat $C$ with $T=0$ information
in Fig.~\ref{fig:heatfrust_moment}. For the specific heat in the
supercritical regime the derivative $dC/dT$ is set to zero, too. 

Obviously, frustration is
favourable for the range of applicability of the $1/T$ expansion.
Without frustration we had to include much higher orders to achieve
similar agreement down to  $T/J\approx 0.25$. The
Figs.~\ref{fig:chifrust_moment} and \ref{fig:heatfrust_moment}
depict data for a specific value of frustration. But the raw
data as given in Appendix A2 allow the calculation of susceptibilities and
specific heats for any value of frustration.

Three possible sources for the improvement of the $1/T$ description
by frustration are conceivable. One is the appearance of a gap due
to frustration. But for $\alpha=\alpha_c$ we found qualitatively the
same behaviour so that this explanation can be excluded.
A second idea concerns the dominance of logarithmic corrections.
Since our ans\"atze are not fit to represent these corrections the
agreement must deteriorate once logarithmic corrections become important
on lowering the temperature. If this mechanism were the dominant one one
should expect a significantly improved agreement at the critical frustration.
The actual comparison (not shown), however, does not display a
significantly improved agreement. So the logarithmic corrections seem
to be not the main problem of a correct representation \cite{note-ab}.

The third possible explanation is a reduction of the spin wave
velocity or, put differently, of the whole dispersion. 
Analytically, it is known in leading
order of an expansion around the dimer limit
that frustration lowers the mobility of the excitation 
\cite{uhrig96b}. Numerical 
results show the same, see Eq.~(\ref{eq:chi0frust}). Indeed,
the positions of the maxima and the lower bound of the range of validity
scale roughly like the spin wave velocity as given by Eq.~(\ref{eq:chi0frust}).
Hence, our results indicate that the estimate (\ref{eq:estimate}) is
valid to the extent that it establishes a proportionality 
$T_{\rm min} \propto v_S/n$.

Summarising this section we state that the Pad\'e
representations with low temperature information incorporated show
very good agreement down to rather low values of $T/J$. 
In particular the maxima of the physical
quantities susceptibility and specific heat
are sufficiently well described. With the full dependence of
the model parameter one has a powerful tool to fit the parameters
to experimental data in a very fast and convenient way.

\section{Representation with Dispersion Data}
\label{sec:dispers}

In this section a different kind of representation is illustrated.
It is also based on the idea to incorporate low temperature
information in a high temperature expansion. As in the previous
section, the approach is motivated by 
approximations for the susceptibility of a dimerised and frustrated
$S=\frac{1}{2}$ chain in Refs.~\cite{troye94,uhrig98c}. There
the authors approximate the magnetic susceptibility by using an exclusion
statistics appropriate for excited triplets  in the dimer model.
To first approximation these excitations are often treated as free bosons.
Yet it is obvious that they are hard-core bosons since there cannot be more
than one on each dimer. This basic fact was included in Ref.~\cite{troye94}.
In Ref.~\cite{uhrig98c} the interaction beyond the hard-core exclusion
was incorporated on a mean-field level.

Defining the partition function
\begin{equation}
  \label{eq:partition}
  z(\beta)=\frac{1}{2\pi}\int_{-\pi}^{\pi}d\!k e^{-\beta\omega(k)}\ .
\end{equation}
for a single excitation one obtains  the hard-core
\begin{equation}
  \label{eq:dispchi0}
  \chi_0 = \beta \frac{z(\beta)}{1+3z(\beta)}
\end{equation}
for the hard-core exclusion statistics. On the mean-field level
one obtains
\begin{equation}
  \label{eq:dispchi1}
  \chi = \frac{\chi_0}{1+J_{\rm eff}\chi_0}
\end{equation}
where $J_{\rm eff}$ can be determined either in the limit of
strong dimerisation or in such a way that the Curie-Weiss constant
is correct. Both methods do not differ much \cite{uhrig98c}.
The approach (\ref{eq:dispchi1}) is very successful in describing
triplets with small dispersion \cite{luthi00a} as in SrCu$_2$(BO$_3$)$_2$
\cite{kagey00a}.

Formulae (\ref{eq:dispchi0},\ref{eq:dispchi1}) suggest to represent 
$T\chi$ essentially as function of $z(\beta)$. In
the lowest order the representation should reproduce
Eq.~(\ref{eq:dispchi0}). Furthermore, it should allow to incorporate the 
information of the high temperature expansion in a natural way 
and the ansatz should be as simple
as possible. Our choice is
\begin{equation}
  \label{eq:Tchiguess}
  T\chi = \cfrac{c_0 z(\beta)}{1 +
      \cfrac{c_1v(\beta)}{1+
        \cfrac{ c_2v(\beta)}{1+c_3 v(\beta) \cdots}}}
\end{equation}
with
\begin{equation}
  \label{eq:v}
  v(\beta) = 1 - z(\beta)\ .
\end{equation}
The variable $v(\beta)$ is chosen such that $v(\beta) \propto \beta$
for $\beta\to0$ so that the coefficients in (\ref{eq:Tchiguess}) 
can be determined straightforwardly from the
 high temperature expansion.  Of course, the choice (\ref{eq:Tchiguess})
is just one of many possible choices so that a certain degree of 
arbitrariness remains. We tried also other choices, for instance an ansatz
extending (\ref{eq:dispchi1})
where $\chi_0$ is taken as variable instead of $v$. Our observation is
that the particular choice does not matter much so that we present here
the easiest ansatz we could think of.

The low temperature information incorporated in (\ref{eq:Tchiguess}) 
is in the dispersion relation $\omega(k)$. In order to demonstrate
the approach we apply it to the unfrustrated chain where we can rely
on exact results for the dispersion \cite{cloiz62}
\begin{equation}
  \label{eq:omega}
  \omega(k) = \frac{\pi}{2}\sin (k)\ .
\end{equation}
We are aware that the unfrustrated case is not particularly
suited for the approach (\ref{eq:Tchiguess}) as motivated above.
The elementary excitations are $S=1/2$ spinons\cite{fadde81}, not magnons.
In this respect, we are choosing a difficult test case for which we will show
 that the approach works very well. On the other hand, it is
known that the main  weight of the  dynamic structure factor 
\cite{mulle81,karba97} is
located close to the lower boundary given by (\ref{eq:omega}) so that
the use of (\ref{eq:omega}) as ``magnon dispersion'' is justifiable. 

Evaluation of the integral (\ref{eq:partition}) yields
\begin{equation}
  z(\beta) = I_0\left(\frac{1}{2}\beta\pi\right) -
   L_0\left(\frac{1}{2}\beta\pi \right)
  \label{eq:partitioneval}
\end{equation}
with the modified Bessel function of the first kind $I_{\nu}$ and
the modified Struve function $L_{\nu}$ as defined in
Ref.~\cite{abram64}.

By construction, the ansatz (\ref{eq:Tchiguess}) is able to fulfill
the high temperature limit $\beta\to 0 $ where 
$T\chi \to 1/4$. It does so if $c_0$ is set to $1/4$.
It is a very favourable feature that the opposite limit of
vanishing temperature  $\beta\to \infty $ where
$T\chi\to T/\pi^2$ \cite{griff64,yang66b} can also be reproduced.
Using
\begin{mathletters}
\begin{eqnarray}
z(\beta)&=&2/\pi\int_0^{\pi/2}e^{-\beta\pi/2\sin(k)d\!k} \\
 &\underset{\beta\to\infty}{=} & 2/\pi\int_{0}^{\infty} 
e^{-\beta\pi/2 k}d\!k = 4/(\beta\pi^2)
\end{eqnarray}
\end{mathletters}
one easily sees that the correct $T\to0$ limit is obtained if
\begin{equation}
\label{eq:valid}
 1=4c_0/(1+c_1/(1+c_2/(1+c_3/\cdots)))
\end{equation}
holds.

\begin{figure}[t]
  \begin{center}
    \includegraphics[width=\columnwidth]{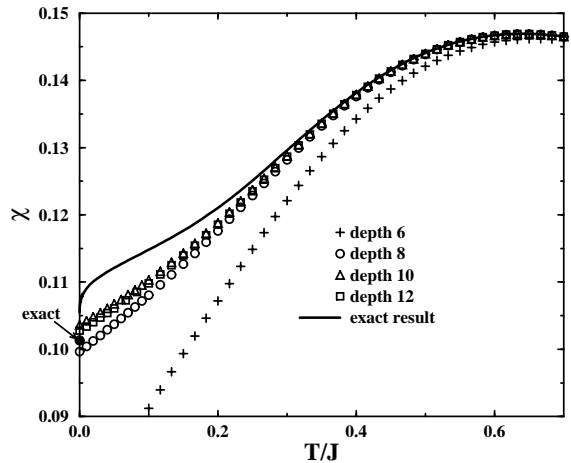}
    \caption{Various orders of continued fraction representations of
      Eq.~(\ref{eq:Tchiguess}) for the magnetic susceptibility of a
      Heisenberg chain in comparison to the exact result obtained by
      Bethe ansatz. The arrow indicates the exact result at $T=0$.}
    \label{fig:dispchiv}
  \end{center}
\end{figure}
Fig.~\ref{fig:dispchiv} shows various continued fraction
representations of the magnetic susceptibility of a Heisenberg
chain. Already low orders of the representation show good
agreement with the exact result down to low $T/J$. Even the
sixth order representation describes position and height of
the maximum fairly well. For order 8 and above, the maximum is
perfectly described. Even the value of the zero temperature
susceptibility is very close to its exact value which could
not be expected since the validity of (\ref{eq:valid}) is not built-in.

The 12th order
representation fits very well down to $T/J\approx 0.3$, which is
almost the result of the $[12,12]$-Pad\'e approximant in
Fig.~\ref{fig:chi_linked} obtained from a much costlier series up
to order 24. Orders above 12 are difficult to implement with the
dispersion information
since the determination of the constants $c_i$ becomes very tedious.
Parallely, the improvement obtained becomes smaller and smaller.

Looking at Fig.~\ref{fig:dispchiv} closely 
it can be concluded that only the logarithmic terms in
the susceptibility \cite{lukya98,klump98} spoil the
agreement at low temperatures $T/J< 0.3$. If these terms
are explicitly incorporated the agreement in the whole temperature 
range can be conveniently described \cite{johns00b}.

\begin{figure}[t]
  \begin{center}
    \includegraphics[width=\columnwidth]{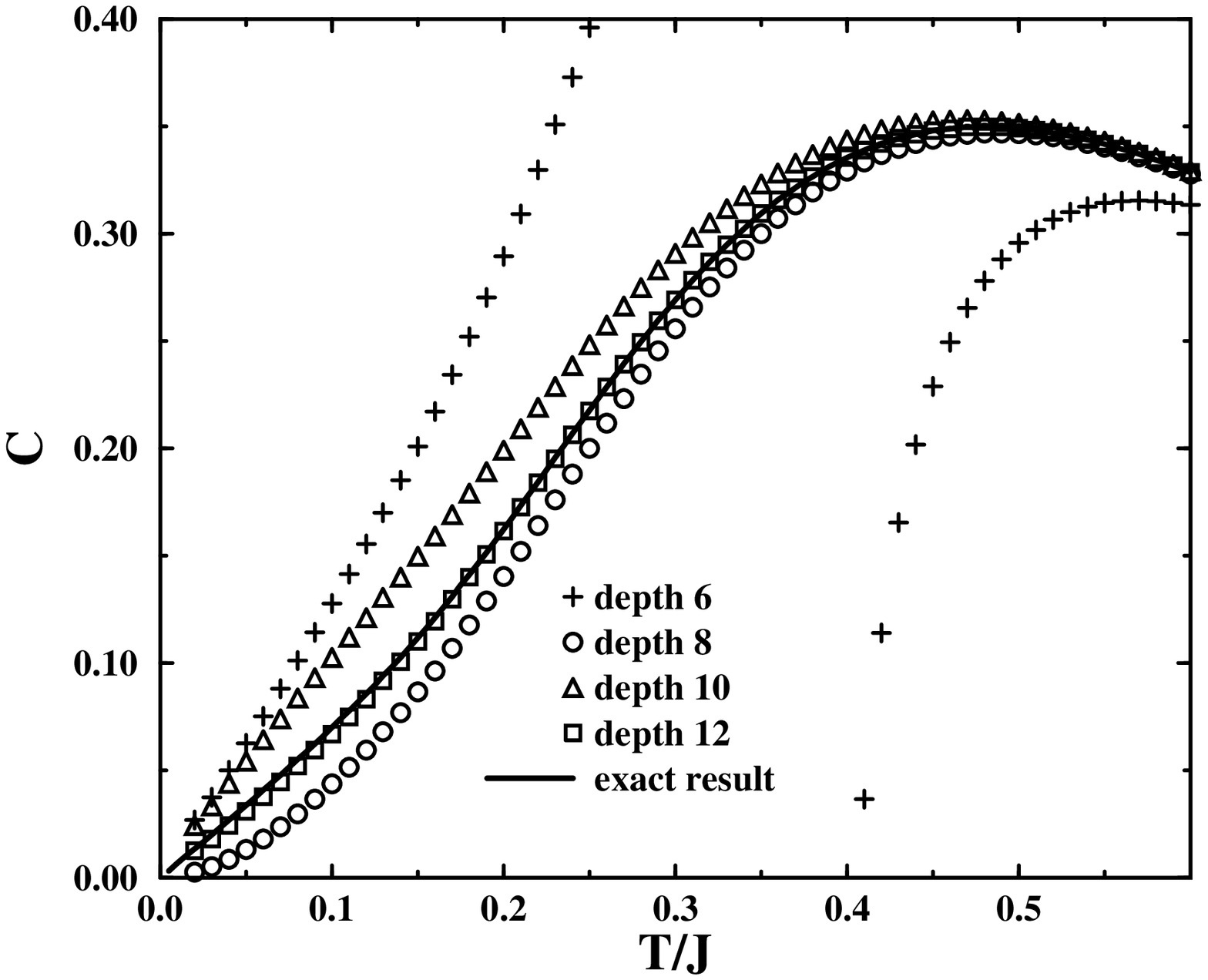}
    \caption{Various orders of continued fraction representations of
      Eq.~(\ref{eq:Tcguess}) for the specific heat of a
      Heisenberg chain in comparison to the exact result obtained by
      Bethe ansatz.}
    \label{fig:dispcv}
  \end{center}
\end{figure}

It is, however, not our aim to provide a fit to a result which
is known analytical. For this reason we do not follow the route
to incorporate the logarithmic terms into the ansatz 
(\ref{eq:Tchiguess}). By the results depicted in 
Fig.~\ref{fig:dispchiv} we have demonstrated that the inclusion of
$T=0$ information in an ansatz of high temperature expansion
improves the range of validity considerably. In particular,
already a small number of high temperature coefficients allows
a satisfyingly accurate description of the overall form of the
physical quantity under study. To corroborate this conclusion
we present in Fig.~\ref{fig:dispcv} the analogous result for
the specific heat. It is based on the ansatz
\begin{equation}
  \label{eq:Tcguess}
  C = \frac{3}{2}\beta^2 \cfrac{d_0 \left(z''-3(z')^2/(1+3z)\right)}{1 +
      \cfrac{d_1v(\beta)}{1+
        \cfrac{ d_2v(\beta)}{1+d_3 v(\beta) \cdots}}}
\end{equation}
where $z'$ and $z''$ stand for the first and the second derivative
of $z$ with respect to $\beta$, respectively. The ansatz
(\ref{eq:Tcguess}) is motivated by the result
\begin{equation}
  \label{eq:Tcttw}
  C = \frac{3}{2}\beta^2 \cfrac{ \left(z''-3(z')^2/(1+3z)\right)}{1+3z}
\end{equation}
derived from the free energy including exclusion statistics \cite{troye94}.
As for the susceptibility the agreement between approximate ansatz
and exact results is good even for low depths of the 
continued fraction (\ref{eq:Tcguess}). 
Note, however, the spurious pole at about $T=0.3 J$
occuring in the representation of
depth 6. This phenomenon cannot be excluded in 
 Pad\'e representations so that one should always consider various
depths in order to judge which features are meaningful.
The 12th order result agrees 
excellently with the exact result which shows the efficiency of the
ansatz (\ref{eq:Tcguess}).

Against the ans\"atze (\ref{eq:Tchiguess},\ref{eq:Tcguess})
one may object that the dispersion $\omega(k)$ will not be available
in general. Of course, exact results for the dispersion are as rare
as exact results for susceptibilities or specific heats. But there are
a number of good approximate
methods which provide $\omega(k)$ at zero temperature,
for instance perturbative expansions \cite{knett00a}. Their results can
be taken to refine and to supplement the high temperature expansions.
Or it can be reasonable to use the experimentally determined results
for $\omega(k)$ in order to understand the thermodynamic quantities.

The fact that already low orders of a high temperature expansion
can be sufficient to provide good estimates for $\chi(T)$ and $C(T)$
is especially important for higher dimensional models in $d=2$ or $d=3$.
In these cases higher orders cannot be obtained due to the quickly
rising number of sites or number of clusters 
to be considered.

\section{Summary}
\label{sec:conclusion} 
The main objective of the present paper is to illustrate a general way
to obtain very fast and convenient analytical formulae for
standard thermodynamical quantities such as the magnetic susceptibility
and the specific heat. 
 So the results should be viewed mainly as effective tools
for quick data analysis.
The route to such formulae comprises two steps.
The first one is a high
temperature expansion in $\beta$ performed symbolically on computers 
providing the coefficients in analytic form.
The second one is the use of an optimised representation of
the results. In this second step the inclusion of additional
information at zero temperature available from
other sources is particularly useful.
 It is not our objective to compute by the methods presented
unkown low temperature physics.

For the sake of illustration we considered in the present article
frustrated Heisenberg chains, i.e. chains of $S=1/2$ spins with
nearest and next-nearest neighbour couplings $J$ and $\alpha J$,
respectively. For these chains we provided the high temperature
coefficients for the  magnetic susceptibility
and the specific heat up to order 10 in the frustrated case and
up to order 24 in the unfrustrated case.

For topologically simple lattices the linked cluster approach is
the most efficient since it avoids to compute powers of the hamiltonian
on unnecessarily large systems. The actual expansion is done for subsystems,
the so-called clusters. The price to pay is the necessary 
sophisticated bookkeeping of the clusters.

In order to be able to treat straightforwardly also more complicated lattices
(or more  complicated topologies of couplings such as frustrating
couplings) we abandoned the calculation on subsystems in the moment algorithm.
In order not to be overwhelmed by
the quickly rising dimension of the Hilbert space we identified
the original trace with an expectation value on an extended Hilbert
space. This trick reduces the number of terms in the application
of the hamiltonian from $L 2^N$ to $L$ where $L$ is the number of
bonds and $N$ the number of sites. In the moment algorithm the
inclusion of frustration is not more complicated than the inclusion
of any other additional coupling. This is in contrast to quantum
Monte Carlo approaches where frustration leads generically
to the severe sign problem.

For not too low temperatures a very good agreement could be achieved.
Including further information on the low temperature the agreement
can be improved further. In particular, the use of information
on the zero temperature dispersion turned out to be very efficient.
In this way, even relatively low orders allow a good description of
the thermodynamic quantities under study. The necessary dispersion
information can be taken from exact or approximate results.
Taking experimental dispersion results allows to check the consistency
of the model assumed. 

Due to the possibility to reach satisfactory results already in
low orders of the high temperature expansion the application  to
higher dimensional cases such as the strongly frustrated 
Shastry-Sutherland model \cite{kagey99,weiho99a} is possible and
will be reported elsewhere. Work on gapped systems such as dimerised
chains is in progress.

\section{Acknowledgements}
We acknowledge many useful discussions with E.~M\"uller-Hartmann
and H.~Monien. We are indebted to A.~Kl\"umper for the Bethe ansatz data and
to R.~Raupach and F.~Sch\"onfeld for the DMRG results which we
used as benchmarks. Equally, helpful remarks by A.~Honecker are acknowledged.
 This work was supported by the Deutsche
Forschungsgemeinschaft in the SFB 341 and in the Schwer\-punkt
1073. The computations were mainly done on machines of the
Regional Computing Center of the University of Cologne.

\onecolumn

\appendix
\label{appendix}
\section{Coefficients}
\subsection{Unfrustrated chain, moment algorithm}
Here the results for the unfrustrated chain are presented. The
susceptibility and the specific heat have been computed up to order 16.

\setlength{\unitlength}{\linewidth}
\begin{picture}(1,0.47)
\put(0,0){
\begin{minipage}[b]{0.46\linewidth}
\subsubsection{Susceptibility \label{sec:momentsusiNN}}

\renewcommand{\arraystretch}{1.5} 
\renewcommand{\arrayrulewidth}{0.4pt} 
\renewcommand{\doublerulesep}{0pt} 
\begin{center}

\begin{tabular}{|c|>{$}c<{$} !{\vrule width 1pt} c|>{$}c<{$}|} \hline
  $n$ & a_n & $n$ & a_n  \\ \hline \hline \hline
  \phantom{0}0\phantom{0} &\phantom{000000}\frac{1}{4}\phantom{000000}
           & \phantom{0}9\phantom{0}  &\frac{3737}{74317824} \\ \hline
  1 &-\frac{1}{8} & 10 &-\frac{339691}{5945425920}  \\ \hline
  2 & $\footnotesize{0}$  & 11 &-\frac{1428209}{54499737600}  \\ \hline
  3 &\frac{1}{96} & 12 &\frac{18710029}{2242274918400}  \\ \hline
  4 &\frac{5}{1536} & 13 &\frac{7045849}{809710387200}  \\ \hline
  5 &-\frac{7}{5120} & 14 &-\frac{358847}{3957275492352}  \\ \hline
  6 &-\frac{133}{122880} 
& 15 &-\frac{65174099663}{28566582460416000}  \\ \hline
  7 &\frac{1}{16128} & 16 &-\frac{258645079463}{498616712036352000}  \\ \hline
  8 &\frac{1269}{4587520} &    &  \\ \hline
\end{tabular}

\end{center}
Series coefficients $a_{n}$ for the high temperature  expansion of
the
  magnetic susceptibility $\chi = \frac{1}{T}\sum_{n}a_n(\beta J)^n$.

\end{minipage}
}

\put(0.5,0){
\begin{minipage}[b]{0.46\linewidth}
\subsubsection{Specific heat \label{sec:momentcNN}}

\renewcommand{\arraystretch}{1.5} 
\renewcommand{\arrayrulewidth}{0.4pt} 
\renewcommand{\doublerulesep}{0pt} 
\begin{center}
\begin{tabular}{|c|>{$}c<{$} !{\vrule width 1pt} c|>{$}c<{$}|} \hline
  $n$ & a_n & $n$ & a_n  \\ \hline \hline \hline
  \phantom{0}0\phantom{0} &\phantom{000000}$\footnotesize{0}$\phantom{0000000}
           & \phantom{0}9\phantom{0}  &-\frac{4303}{688128} \\ \hline
  1 &\phantom{000000}$\footnotesize{0}$\phantom{0000000} &
                      10 &-\frac{334433}{110100480}  \\ \hline
  2 & \frac{3}{16}  & 11 &\frac{37543}{31457280}  \\ \hline
  3 &\frac{3}{32} & 12 &\frac{3987607}{3170893824}  \\ \hline
  4 &-\frac{15}{256} & 13 &-\frac{1925339}{41523609600}  \\ \hline
  5 &-\frac{15}{256} & 14 &-\frac{369233453}{930128855040}  \\ \hline
  6 &\frac{21}{4096} & 15 &-\frac{31504270817}{362750253465600}  \\ \hline
  7 &\frac{917}{40960} & 16 &\frac{851758334701}{8706006083174400}  \\ \hline
  8 &\frac{1417}{327680} &    &  \\ \hline
\end{tabular}
\end{center}
Series coefficients $a_{n}$ for the high temperature expansion of
the
  magnetic specific heat $C = \sum_{n}a_n(\beta J)^n$.

\end{minipage}
}

\end{picture}

\subsection{Frustrated chain, moment algorithm}
The coefficients for the results of frustrated chain are
presented. The magnetic susceptibility and the magnetic specific heat 
are expanded up to order 10
in $\beta J$.

\subsubsection{Susceptibility \label{sec:momentsusiNNN}}

\renewcommand{\arraystretch}{1.5} 
\renewcommand{\arrayrulewidth}{0.4pt} 
\renewcommand{\doublerulesep}{0pt} 
\begin{center}
\begin{tabular}{|>{\centering}p{0.056\linewidth}|%
    >{$} c <{$}!{\vrule width 1pt}
    >{\centering} p{0.056\linewidth}|%
    >{$}c <{$}!{\vrule width 1pt}
    >{\centering} p{0.056\linewidth}|%
    >{$}c <{$}!{\vrule width 1pt}
    >{\centering} p{0.056\linewidth}|%
    >{$}c <{$}!{\vrule width 1pt}
    >{\centering} p{0.056\linewidth}|%
    >{$}c <{$}!{\vrule width 1pt}
    >{\centering} p{0.056\linewidth}|%
    >{$}c <{$}|}%
     \hline
  $(n,k)$&a_{n,k}&$(n,k)$&a_{n,k}&$(n,k)$&a_{n,k}&$(n,k)$&a_{n,k}
    &$(n,k)$&a_{n,k}&$(n,k)$&a_{n,k}   \\ \hline \hline \hline
  (0,0)&\frac{1}{4}&(4,1)&-\frac{23}{768}
    &(6,1)&\frac{9}{1280}&(7,5)&\frac{943}{368640}
    &(8,8)&\frac{1269}{4587520}&(10,0)&-\frac{339691}{5945425920} \\
    \cline{1-2} \cline{9-10}
  (1,0)&-\frac{1}{8}&(4,2)&\frac{1}{512}
    &(6,2)&\frac{221}{61440}&(7,6)&\frac{67}{368640}
    &(9,0)&\frac{3737}{74317824}&(10,1)&-\frac{22843}{1486356480} \\
  (1,1)&-\frac{1}{8}&(4,3)&-\frac{1}{96}
    &(6,3)&-\frac{163}{92160}&(7,7)&\frac{1}{16128}
    &(9,1)&-\frac{34337}{23592960}&(10,2)&\frac{15205963}{5945425920} \\
    \cline{1-2} \cline{7-8}
  (2,0)&$\footnotesize{0}$&(4,4)&\frac{5}{1536}
    &(6,4)&\frac{7}{15360}&(8,0)&\frac{1269}{4587520}
    &(9,2)&\frac{14125}{4128768}&(10,3)&-\frac{311903}{82575360} \\
    \cline{3-4}
  (2,1)&\frac{1}{8}&(5,0)&-\frac{7}{5120}
    &(6,5)&\frac{23}{7680}&(8,1)&-\frac{23629}{20643840}
    &(9,3)&-\frac{1249}{35389440}&(10,4)&\frac{9659}{3932160} \\
  (2,2)&$\footnotesize{0}$&(5,1)&-\frac{49}{6144}
    &(6,6)&-\frac{133}{122880}&(8,2)&-\frac{58651}{13762560}
    &(9,4)&\frac{317}{229376}&(10,5)&-\frac{1177787}{825753600} \\
    \cline{1-2} \cline{5-6}
  (3,0)&\frac{1}{96}&(5,2)&\frac{37}{1536}
    &(7,0)&\frac{1}{16128}&(8,3)&\frac{28751}{5160960}
    &(9,5)&-\frac{969}{655360}&(10,6)&\frac{599639}{594542592} \\
  (3,1)&\frac{1}{128}&(5,3)&-\frac{1}{128}
    &(7,1)&\frac{5863}{1474560}&(8,4)&-\frac{59}{20160}
    &(9,6)&\frac{93463}{61931520}&(10,7)&\frac{791221}{1486356480} \\
  (3,2)&-\frac{1}{32}&(5,4)&\frac{1}{512}
    &(7,2)&-\frac{805}{73728}&(8,5)&-\frac{877}{1290240}
    &(9,7)&-\frac{67097}{82575360}&(10,8)&-\frac{367481}{1486356480} \\
  (3,3)&\frac{1}{96}&(5,5)&-\frac{7}{5120}
    &(7,3)&\frac{3023}{737280}&(8,6)&\frac{5389}{20643840}
    &(9,8)&-\frac{361}{1720320}&(10,9)&\frac{22433}{148635648} \\
  \cline{1-2} \cline{3-4}
  (4,0)&$\phantom{0}$\frac{5}{1536}$\phantom{0}$&(6,0)&
  $\phantom{0}$-\frac{133}{122880}$\phantom{0}$
    &(7,4)&$\phantom{0}$-\frac{381}{81920}$\phantom{0}$ &(8,7)&
  $\phantom{0}$-\frac{1271}{1720320} $\phantom{0}$
    &(9,9)&$\phantom{0}$\frac{3737}{74317824}&(10,10)$\phantom{0}$&
  $\phantom{0}$ -\frac{339691}{5945425920}$\phantom{0}$  \\ \hline
\end{tabular}
\end{center}
Series coefficients $a_{n,k}$ for the high temperature expansion
of the
  magnetic susceptibility of the frustrated chain
  $\chi =\frac{1}{T}\sum_{n,k}a_{n,k}\alpha^k(\beta J)^n$.

\subsubsection{Specific heat \label{sec:momentcNNN}}

\renewcommand{\arraystretch}{1.5} 
\renewcommand{\arrayrulewidth}{0.4pt} 
\renewcommand{\doublerulesep}{0pt} 
\begin{center}
\begin{tabular}{|>{\centering}p{0.056\linewidth}|%
    >{$} c <{$}!{\vrule width 1pt}
    >{\centering} p{0.056\linewidth}|%
    >{$}c <{$}!{\vrule width 1pt}
    >{\centering} p{0.056\linewidth}|%
    >{$}c <{$}!{\vrule width 1pt}
    >{\centering} p{0.056\linewidth}|%
    >{$}c <{$}!{\vrule width 1pt}
    >{\centering} p{0.056\linewidth}|%
    >{$}c <{$}!{\vrule width 1pt}
    >{\centering} p{0.056\linewidth}|%
    >{$}c <{$}|}%
     \hline
  $(n,k)$&a_{n,k}&$(n,k)$&a_{n,k}&$(n,k)$&a_{n,k}&$(n,k)$&a_{n,k}
    &$(n,k)$&a_{n,k}&$(n,k)$&a_{n,k}   \\ \hline \hline \hline
  (0,0)&$\footnotesize{0}$&(4,1)&-\frac{3}{32}
    &(6,1)&\frac{63}{512}&(7,5)&-\frac{245}{8192}
    &(8,8)&\frac{1417}{327680}&(10,0)&-\frac{334433}{110100480} \\
    \cline{1-2} \cline{9-10}
  (1,0)&$\footnotesize{0}$&(4,2)&-\frac{3}{32}
    &(6,2)&-\frac{363}{4096}&(7,6)&$\footnotesize{0}$
    &(9,0)&-\frac{4303}{688128}&(10,1)&\frac{92629}{2752512} \\
  (1,1)&$\footnotesize{0}$&(4,3)&$\footnotesize{0}$
    &(6,3)&\frac{17}{512}&(7,7)&\frac{917}{40960}
    &(9,1)&\frac{2613}{573440}&(10,2)&-\frac{420475}{11010048} \\
    \cline{1-2} \cline{7-8}
  (2,0)&\frac{3}{16}&(4,4)&-\frac{15}{256}
    &(6,4)&\frac{105}{1024}&(8,0)&\frac{1417}{327680}
    &(9,2)&\frac{3855}{57344}&(10,3)&\frac{59305}{2752512} \\
    \cline{3-4}
  (2,1)&$\footnotesize{0}$&(5,0)&-\frac{15}{256}
    &(6,5)&$\footnotesize{0}$&(8,1)&-\frac{4793}{61440}
    &(9,3)&\frac{1}{10240}&(10,4)&-\frac{138811}{2752512} \\
  (2,2)&\frac{3}{16}&(5,1)&\frac{25}{128}
    &(6,6)&\frac{21}{4096}&(8,2)&\frac{2323}{24576}
    &(9,4)&-\frac{261}{286720}&(10,5)&\frac{51701}{1376256} \\
    \cline{1-2} \cline{5-6}
  (3,0)&\frac{3}{32}&(5,2)&-\frac{5}{128}
    &(7,0)&\frac{917}{40960}&(8,3)&-\frac{59}{960}
    &(9,5)&\frac{5901}{81920}&(10,6)&\frac{27641}{2752512} \\
  (3,1)&-\frac{9}{32}&(5,3)&\frac{15}{128}
    &(7,1)&-\frac{2611}{40960}&(8,4)&\frac{35}{2048}
    &(9,6)&-\frac{2411}{143360}&(10,7)&-\frac{1817}{917504} \\
  (3,2)&$\footnotesize{0}$&(5,4)&$\footnotesize{0}$
    &(7,2)&-\frac{119}{4096}&(8,5)
    &$\phantom{\footnotesize{0}}$-\frac{407}{61440}$\phantom{\footnotesize{0}}$
    &(9,7)&-\frac{2229}{573440}&(10,8)&\frac{38993}{1572864} \\
  (3,3)&\frac{3}{32}&(5,5)&-\frac{15}{256}
    &(7,3)&-\frac{413}{4096}&(8,6)&-\frac{2449}{40960}
    &(9,8)&$\footnotesize{0}$&(10,9)&$\footnotesize{0}$ \\ \cline{1-2}\cline{3-4}
  (4,0)&$\phantom{\footnotesize{0}}$-\frac{15}{256}$\phantom{\footnotesize{0}}$&(6,0)&
       $\phantom{\footnotesize{0}}$ \frac{21}{4096}$\phantom{\footnotesize{0}}$
    &(7,4)&$\phantom{\footnotesize{0}}$\frac{651}{20480}$\phantom{\footnotesize{0}}$
    &(8,7)&$\footnotesize{0}$
    &(9,9)&$\phantom{\footnotesize{0}}$-\frac{4303}{688128}$\phantom{\footnotesize{0}}$
    &(10,10)&$\phantom{\footnotesize{0}}$
    -\frac{334433}{110100480}$\phantom{\footnotesize{0}}$
    \\ \hline
\end{tabular}

\end{center}
Series coefficients $a_{n,k}$ for the high temperature expansion
of the
  magnetic specific heat of the frustrated chain $C =
  \sum_{n,k}a_{n,k}\alpha^k(\beta J)^n$. 

\subsection{Unfrustrated chain, linked cluster expansion \label{sec:linkedcoef}}

\subsubsection{Susceptibility \label{sec:linkedsusi}}

\renewcommand{\arraystretch}{1.5} 
\renewcommand{\arrayrulewidth}{0.4pt} 
\renewcommand{\doublerulesep}{0pt} 
\begin{center}
\begin{tabular}{|c|%
    >{\footnotesize} c !{\vrule width 1pt}
    c|>{\footnotesize}c!{\vrule width 1pt}
    c|>{\footnotesize}c!{\vrule width 1pt}
    c|>{\footnotesize}c!{\vrule width 1pt}
    c|>{\footnotesize}c|}   \hline
  $n$&\normalsize{$a_n$}&$n$&\normalsize{$a_n$}&$n$&\normalsize{$a_n$}&$n$&
   \normalsize{$a_n$}&$n$&\normalsize{$a_n$} \\ \hline \hline \hline
  0&1.0 &5&-4032.0
    &10&-9565698560.0 &15&-205019990184689664.0
    &20&-18366266410738921187573760.0 \\ \hline
  1&-4.0 &6&-89376.0
    &11&-210597986304.0 &16& -3169755454477500416.0
    &21&-40780317289246872850923520.0 \\ \hline
  2&0.0 &7& 163840.0
    &12&3486950684672.0  &17&208763541109969256448.0
    &22&38668138493195891009425244160.0\\ \hline
  3&64.0 &8& 26313984.0
    &13&203634731188224.0 &18& 8342101010835559022592.0
    &23& 983734184997038611238624428032.0\\ \hline
  4& 400.0 &9&191334400.0
    &14&-127324657152000.0 &19& -175912858271144581529600.0
    &24&-75650797544886562610211717119286.7   \\  \hline
\end{tabular}
\end{center}
Series coefficients for the linked cluster expansion of the
magnetic
  susceptibility for the Heisenberg chain with $\chi =
  \frac{1}{4T} \sum_n \frac{a_n}{(n+1)!} ( \frac{J}{4T} )^n $. 

\subsubsection{Specific heat \label{sec:linkedheat}}

\renewcommand{\arraystretch}{1.5} 
\renewcommand{\arrayrulewidth}{0.4pt} 
\renewcommand{\doublerulesep}{0pt} 
\begin{center}
\begin{tabular}{|c|%
    >{\footnotesize} c !{\vrule width 1pt}
    c|>{\footnotesize}c!{\vrule width 1pt}
    c|>{\footnotesize}c!{\vrule width 1pt}
    c|>{\footnotesize}c!{\vrule width 1pt}
    c|>{\footnotesize}c|}   \hline
  $n$&\normalsize{$a_n$}&$n$&\normalsize{$a_n$}&$n$&\normalsize{$a_n$}&$n$&
   \normalsize{$a_n$}&$n$&\normalsize{$a_n$} \\ \hline \hline \hline
  0&0.0 &5&-7200.0
    &10& -11558004480.0&15&-121944211136778240.0
    &20& 96147483542540314214400.0\\ \hline
  1& 0.0 &6&15120.0
    &11&199812856320.0 &16& 8791781390116945920.0
    &21&1279121513829538179364945920.0\\ \hline
  2& 6.0&7& 1848672.0
    &12& 10106191180800.0 &17&310402124957945954304.0
    &22&27962069861743501862336200704.0\\ \hline
  3&36.0 &8&11426688.0
    &13&-19376365252608.0 &18&-7225535925744106143744.0
    &23&-2398518627113966015427501883392.0\\ \hline
  4& -360.0 &9&-594846720.0
    &14&-9289795522775040.0&19&-643407197363813620776960.0
    &24& -129834725539335848980192847460554.1120  \\  \hline
\end{tabular}
\end{center}
Series coefficients for the linked cluster expansion of the
magnetic
  specific heat for the Heisenberg chain with $C =
   \sum_n \frac{a_n}{n!} ( \frac{J}{4T} )^n $. 


\begin{thebibliography}{10}

\bibitem{johns00a}
D.~C. Johnston {\it et~al.},   cond-mat/0001147  (2000).

\bibitem{fabri97a}
K. Fabricius, U. L\"ow, and J. Stolze, Phys. Rev. B {\bf 55},  5833  (1997).

\bibitem{bursi95}
R. Bursill {\it et~al.}, J. Phys.: Condens. Matter {\bf 7},  8605  (1995).

\bibitem{shiba97}
N. Shibata, J. Phys. Soc. Jpn. {\bf 66},  2221  (1997).

\bibitem{wang98}
X.~Q. Wang and T. Xiang, Phys. Rev. B {\bf 56},  5061  (1998).

\bibitem{johns00b}
D.~C. Johnston {\it et~al.},   cond-mat/0003271  (2000).

\bibitem{he90}
H.-X. He, C.~J. Hamer, and J. Oitmaa, J. Phys. A: Math. Gen.
 {\bf 23},  1775  (1990).

\bibitem{gelfa90}
M.~P. Gelfand, R.~R.~P. Singh, and D.~A. Huse, J. Stat. Phys. {\bf 59},  1093
  (1990).

\bibitem{baker64}
G.~A. Baker, G.~S. Rushbrooke, and H.~E. Gilbert, Phys. Rev. {\bf 135},  A1272
  (1964).

\bibitem{oboka67}
T. Obokata, I. Ono, and T. Oguchi, J. Phys. Soc. Jpn. {\bf 23},  516  (1967).

\bibitem{klump93b}
A. Kl\"umper, Z. Phys. B {\bf 91},  507  (1993).

\bibitem{klump98}
A. Kl\"umper, Eur. Phys. J. B {\bf 5},  677  (1998).

\bibitem{klump99a}
A. Kl\"umper, R. Raupach, and F. Sch\"onfeld, Phys. Rev. B {\bf 59},  3612
  (1999).

\bibitem{mulle81}
G. M\"uller, H. Thomas, H. Beck, and J.~C. Bonner, Phys. Rev. B {\bf 24},  1429
   (1981).

\bibitem{cloiz62}
J. des Cloizeaux and J.~J. Pearson, Phys. Rev. {\bf 128},  2131  (1962).

\bibitem{fledd97}
A. Fledderjohann and C. Gros, Europhys. Lett. {\bf 37},  189  (1997).

\bibitem{egger96}
S. Eggert, Phys. Rev. B {\bf 54},  R9612  (1996).

\bibitem{shast81a}
B.~S. Shastry and B. Sutherland, Phys. Rev. Lett. {\bf 47},  964  (1981).

\bibitem{affle97}
I. Affleck,  in {\em Dynamical Properties of Unconventional Magnetic Systems},
edited by A. T. Skjeltorp and D. Sherrington
  (Kluwer Academic Publishers, 1997).

\bibitem{uhrig99a}
G.~S. Uhrig, F. Sch\"onfeld, M. Laukamp, and E. Dagotto, Eur. Phys. J. B {\bf
  7},  67  (1999).

\bibitem{hase93a}
M. Hase, I. Terasaki, and K. Uchinokura, Phys. Rev. Lett. {\bf 70},  3651
  (1993).

\bibitem{bouch96}
J.~P. Boucher and L.~P. Regnault, J. Phys. I France {\bf 6},  1939  (1996).

\bibitem{riera95}
J. Riera and A. Dobry, Phys. Rev. B {\bf 51},  16098  (1995).

\bibitem{casti95}
G. Castilla, S. Chakravarty, and V.~J. Emery, Phys. Rev. Lett. {\bf 75},  1823
  (1995).

\bibitem{fabri98a}
K. Fabricius {\it et~al.}, Phys. Rev. B {\bf 57},  1102  (1998).

\bibitem{knett00b}
C. Knetter and G.~S. Uhrig, in preparation.

\bibitem{note-ab}
To be cautious, we re-state that statements on the validity of $1/T$ results
depend on the representation used.

\bibitem{uhrig96b}
G.~S. Uhrig and H.~J. Schulz, Phys. Rev. B {\bf 54},  R9624  (1996);
{\it Err.} {\bf 58},  2900  (1998).

\bibitem{troye94}
M. Troyer, H. Tsunetsugu, and D. W\"urtz, Phys. Rev. B {\bf 50},  13515
  (1994).

\bibitem{uhrig98c}
G.~S. Uhrig and B. Normand, Phys. Rev. B {\bf 58},  R14705  (1998).

\bibitem{luthi00a}
B. L\"uthi, private communication  (2000).

\bibitem{kagey00a}
H. Kageyama {\it et~al.}, submitted to Phys. Rev. Lett.

\bibitem{fadde81}
L.~D. Faddeev and L.~A. Takhtajan, Phys. Lett. {\bf 85A},  375  (1981).

\bibitem{karba97}
M. Karbach {\it et~al.}, Phys. Rev. B {\bf 55},  12510  (1997).

\bibitem{abram64}
M. Abramowitz and I.~A. Stegun, {\em Handbook of Mathematical Functions} (Dover
  Publisher, New York, 1964).

\bibitem{griff64}
R.~B. Griffiths, Phys. Rev. {\bf 133},  A768  (1964).

\bibitem{yang66b}
C.~N. Yang and C.~P. Yang, Phys. Rev. {\bf 151},  258  (1966).

\bibitem{lukya98}
S. Lukyanov, Nucl. Phys. B {\bf 522},  533  (1998).

\bibitem{knett00a}
C. Knetter and G.~S. Uhrig, Eur. Phys. J. B {\bf 13},  209  (2000).

\bibitem{kagey99}
H. Kageyama {\it et~al.}, Phys. Rev. Lett. {\bf 82},  3168  (1999).

\bibitem{weiho99a}
Z. Weihong, C.~J. Hamer, and J. Oitmaa, Phys. Rev. B {\bf 60},  6608  (1999).

\end{thebibliography}
\end{document}